\documentclass[12pt,a4paper]{article}
\usepackage[utf8]{inputenc}
\usepackage[T1]{fontenc}
\usepackage{times}
\usepackage[margin=2.5cm]{geometry}
\usepackage{natbib}
\usepackage{url}
\usepackage{hyperref}
\usepackage{graphicx}
\usepackage{booktabs}
\usepackage{caption}
\usepackage{subcaption}
\usepackage{xcolor}
\usepackage{microtype}
\usepackage{setspace}

\hypersetup{colorlinks=true, linkcolor=blue!50!black, citecolor=blue!40!black, urlcolor=blue!40!black}
\graphicspath{{figures/}}

\title{Beyond Citations: Comparing Scholarly, Policy, and Patent Impact Across the FT50 Journals}

\author{%
  Arash Hajikhani\textsuperscript{1,2,*}\quad
  Yi Zhang\textsuperscript{3}\quad
  Mengjia Wu\textsuperscript{3}
  \\[0.6em]
  \normalsize\textit{\textsuperscript{1}VTT Technical Research Centre of Finland Ltd, Espoo, Finland}\\
  \normalsize\textit{\textsuperscript{2}LUT University, Lappeenranta, Finland}\\
  \normalsize\textit{\textsuperscript{3}University of Technology Sydney, Sydney, NSW, Australia}\\[0.4em]
  \normalsize\textsuperscript{*}Corresponding author: \texttt{arash.hajikhani@vtt.fi} \quad ORCID: 0000-0003-2032-9180
}
\date{}

\begin{document}
\maketitle

\begin{abstract}
\noindent The Financial Times 50 (FT50) journal list shapes hiring, promotion, accreditation, and research evaluation across business schools worldwide. Yet journals on the list are typically treated as if they represent a homogeneous tier of excellence. We test this assumption by comparing 53 FT50 and recently removed journals across three distinct impact channels: scholarly influence (field-weighted citations and visibility), policy uptake, and technological reach through patent citations. Using a panel of more than 60,000 publications from 2005--2019, we find striking heterogeneity hidden beneath the binary FT50 label. Elite economics journals dominate policy influence, information systems and marketing journals lead technological impact, while many highly cited management journals exhibit limited reach beyond academia. Citation, policy, and patent indicators behave as largely independent dimensions of impact, with a citation-only ranking correlating only moderately with a multidimensional ranking. Nearly half of all journals change quartile once policy and patent indicators are incorporated, demonstrating that assessments based solely on scholarly citations overlook important dimensions of research influence. While the FT50 remains widely used as a binary classification of journal quality, our results reveal a substantial within-list impact spectrum and show that journal rankings are highly sensitive to how impact is defined and measured.
\end{abstract}


\section{Introduction}

The Financial Times 50 list (FT50) is used by deans, accreditation bodies, and tenure committees as a shorthand for journal quality in business and management research. Its constitution and effects have been argued over for two decades. The list was last substantially revised in 2016 and has now, in 2026, been adjusted again with three titles removed and three added \citep{FTRanking2026}. The literature accumulated around the list raises three questions that motivate the present analysis. The first concerns selection: does the FT50 identify the journals where the most consequential research is published, and would alternative bibliometric selections produce a different ordering \citep{Fassin2021,Vidgen2019,Mingers2017}. The second concerns the metric ecology in which FT50 evaluations are made: citation-based indicators have been shown to be vulnerable to volume-driven dilution \citep{Zhang2021} and to the broader publication-volume strain documented across academic publishing \citep{Hanson2024}, to Goodhart-style decay as citation counts have become targets \citep{Fire2019}, and to outright manipulation through paper mills and citation-purchasing services \citep{Ibrahim2025,Meho2025}. The third concerns rigour and relevance: critics argue that the kind of scholarship the list rewards has drifted away from the audiences outside academia that business-school research is supposed to serve \citep{Steingard2026,JackDalal2024}, an argument that becomes harder to evaluate as long as the available evidence is restricted to a single, citation-based dimension of impact.

The most useful empirical question to bring to this debate is comparative. If the FT50 certifies a homogeneous tier of high-quality outlets, then on properly normalized per-publication indicators of citation impact, policy reach, and patent reach the journals on the list should look similar; if the list bundles incommensurable kinds of scholarship, the indicators will spread apart. Existing FT50-focused analyses tend to treat impact as a single number \citep{Zhang2021,Fassin2021} or to focus on one disciplinary slice of the list \citep{Vidgen2019}, and rarely include policy or patent reach. The comparison is provided here on a single panel of all 53 journals currently on or recently removed from the list, using all three channels in parallel, and the resulting comparison is tested against a sensitivity analysis over alternative publication-year windows.

The contribution is descriptive. The aim is not to settle whether the FT50 selects the right journals, but to characterize the within-list quality spectrum on per-publication terms, to identify which journals sit at the top of that spectrum on multiple channels and which sit consistently at the bottom, to test whether the identification is stable across publication-year windows, and to read the 2026 list update against the resulting picture. The motivating question is the one a user of any journal list ought to be able to ask: given two journals on the FT50, which has the larger per-paper reach, and on which kind of impact?

\section{Data and methods}

\subsection{Data sources}

Three coupled data sources are used. The bibliographic backbone is Scopus, accessed through the Elsevier APIs that also underlie SciVal. \citet{Baas2020} document Scopus as a curated bibliometric source, and \citet{Visser2021} compare it with Web of Science, Dimensions, Crossref, and Microsoft Academic; for the social and management sciences that fill the FT50, Scopus is the most complete option for both coverage and citation linking. The indicators built on top of Scopus are computed in SciVal, and the SciVal indicator definitions are followed throughout \citep{ElsevierMetrics2019,ElsevierPatent2019}.\footnote{The two SciVal guidebooks \citep{ElsevierMetrics2019,ElsevierPatent2019} and the 2026 FT list announcement \citep{FTRanking2026} are cited here as documentary references --- vendor methodology documentation and a public list announcement, respectively --- rather than as peer-reviewed sources.} The patent-citation stream in SciVal draws on the five largest patent offices: the European Patent Office, the United States Patent and Trademark Office, the United Kingdom Intellectual Property Office, the Japan Patent Office, and the World Intellectual Property Organization. Patent visibility is constrained by an eighteen-month publication lag and a multi-year examination window, so the patent indicators are most reliable for papers published more than three years before the extract date \citep{ElsevierPatent2019}. Public-policy citations are sourced via Overton, described in detail by \citet{Szomszor2022}, with the strengths and limitations of Overton for research-impact evaluation discussed by \citet{Cristofoletti2025}. Overton has a coverage skew toward English-language and Anglo-American policy producers, and the policy series is treated cautiously throughout.

\subsection{Indicators}
\label{sec:indicators}

Seven per-publication indicators are used; they are directly comparable across journals once volume and field are accounted for. Three measure citation-based scholarly impact. The Field-Weighted Citation Impact (FWCI) is the ratio of citations received by a publication to the world average for similar publications, controlling for subject field, year of publication, and document type \citep{ElsevierMetrics2019,Lariviere2019}; a value of 1.0 corresponds to the world average. The Top-10\% rate is the share of a journal's papers in the global top decile by FWCI within their field-year-document-type cell, computed as Output-in-Top-10\% divided by Scholarly Output. The Field-Weighted Views Impact is the same normalization applied to usage rather than citations. Two measure public-policy reach: policy citations per published paper (Citing Policy Documents divided by Scholarly Output) and the share of papers ever cited by at least one policy document (Scholarly Output cited by Policies divided by Scholarly Output). Two measure industry reach: patent citations per published paper (Patent Citations divided by Scholarly Output) and the share of papers ever cited by at least one patent (Scholarly Output cited by Patents divided by Scholarly Output).

Normalizing every count by Scholarly Output is required by the same logic that motivates FWCI. Aggregate counts of citations, policy mentions, or patent links are mechanical functions of how many papers a journal has published and how long those papers have had to accumulate citations. A journal that publishes 1{,}000 papers per year will accumulate more aggregate citations than one that publishes 50, regardless of per-paper impact. To compare journals, the relevant quantity is the rate per published paper. SciVal already applies this adjustment natively in the patent channel by reporting Patent Citations per Scholarly Output as a denominator-adjusted indicator \citep{ElsevierPatent2019}; the same logic is applied here to the policy channel and to the share of papers reaching either channel.

\subsection{Sample and time window}

All 53 journals tagged FT50 in SciVal as of April 2026 are included. The panel is the union of the pre-2026 FT50 (the 50 journals on the list before the most recent update) and the three additions in 2026 (\textit{Academy of Management Annals}, \textit{American Sociological Review}, \textit{Psychological Science}), with the three removed journals (\textit{Human Relations}, \textit{Journal of Business Ethics}, \textit{Organization Studies}) still present in the panel. Keeping the removed journals in the panel makes the 2026 swap directly observable in the comparative analysis. Three of the listed journals (\textit{Academy of Management Annals}, \textit{Strategic Entrepreneurship Journal}, \textit{Review of Finance}) were founded mid-window; they are kept in the panel and earlier years are left missing.

The headline analysis uses publication years 2005--2019. The lower edge is set by the Overton policy database. \citet{Szomszor2022} document indexed-document growth and citation accrual back to 2000--2005, and 2005 is used here as the earliest year for which per-paper policy reach is reasonably comparable across cohorts; pre-2005 measurements would understate per-paper policy reach simply because the policy documents that would have cited those earlier papers are not all in the database.

The upper edge is set by accrual time, and three separate constraints push it down to 2019 even though the SciVal panel itself runs to 2026. The first is citation accrual. The per-publication citation indicators (Citations per Publication, FWCI, Top-10\% rate, Field-Weighted Views Impact) all need post-publication time for citations and views to accumulate. SciVal's FWCI uses a four-year accrual window, so a paper from 2022 has its FWCI computed on citations received through 2025, which is in principle workable but puts late-window cohorts on a noisier footing than earlier cohorts. The Top-10\% rate is a cumulative-since-publication metric and is more sensitive still: a 2024 paper has had two years to enter the global top decile, while a 2010 paper has had sixteen. The standard scientometric convention, discussed at length in \citet{Lariviere2019} in the context of citation-window length, is to allow at least five years of post-publication accrual when comparing journals on citation-based per-paper rates. Stopping at 2019 gives every paper in the panel between five and twenty years of accrual through the April 2026 extract.

The second constraint is the patent channel, which is tighter than the citation channel. Patents have an eighteen-month publication lag from filing, plus a three- to five-year examination process for grant, plus the time it takes for a granted patent to surface in patent-to-paper citation indexing. The Elsevier Patent Metrics Guidebook \citep{ElsevierPatent2019} notes that patent metrics are most reliable for publication years that predate extraction by at least three years, which for an April 2026 extract means a cutoff at 2022 or earlier. In practice, papers published in 2023, 2024, or 2025 have had little or no exposure to the granted patents that would cite them, regardless of the underlying research. Including those years would push the late-cohort patent rates toward zero mechanically, distorting one of the three channels under measurement.

The third constraint is the policy channel on the right-hand side. Policy uptake of academic research typically lags by several years: a paper has to be published, become known, then be cited in a government report, working paper, or NGO brief that itself has to be indexed by Overton. \citet{Cristofoletti2025} discuss right-censoring of recent policy citations as a known feature of the database. For papers from 2024 or later, Overton has had little time to index the policy documents that would cite them.

Cutting at 2019 gives the panel a window that is reliable on all three channels at once. The 15-year span covers 60{,}705 papers across the 53 journals (median 868 papers per journal, range 170 to 5{,}170), enough for the per-publication ratios to stabilize even for journals with small annual output. The choice of upper edge is itself a parameter, and it is tested directly: every per-journal indicator is replicated under the three alternative windows 2005--2014, 2010--2019, and 2015--2019, and Kendall $\tau$ is reported between the rankings produced under each pair of windows. The 2015--2019 replication sits closest to the right-censored zone and is the most exposed to recency noise, so any drop in stability there gives a measurable bound on how much the upper edge is doing the work.

\subsection{Sub-discipline grouping}

For interpretation the 53 journals are grouped into eleven sub-disciplines, following each journal's primary subject area in the All Science Journal Classification used by Scopus to compute FWCI \citep{Baas2020}, supplemented by the journal-by-journal subfield assignments documented for FT50 journals by \citet{Fassin2021} and the Information Systems clustering used by \citet{Vidgen2019}. The full mapping with sources is given in Appendix~\ref{app:disc}. The grouping is heuristic at the boundaries; the channel-share results are stable under several plausible reassignments.

\subsection{Why a comparative cross-journal analysis is the right framing}
\label{sec:framing}

A natural alternative would be to study how impact at the FT50 has changed over time. That framing was tried first and is reported briefly: the median FWCI across the list declines over the 2005--2023 window, from approximately 3.45 in 2005 to 2.04 in 2020 (Section~\ref{sec:caveats}). Because citations, policy mentions, and patent links all accumulate over time, much of this aggregate decline is mechanical: papers published in 2005 have had more time to be discovered than papers published in 2020. Even FWCI, despite its four-year normalization window, is sensitive to the time-window choices that set the world baseline. The longitudinal trend is therefore mostly informative about citation accrual, not about the relative quality of journals on the list. The substantive question, namely which journals sit at the top of the quality spectrum and which sit at the bottom, is sharpest when journals are compared to each other under the same window and after per-publication normalization. That is the framing of the rest of the paper.

\subsection{Composite ranking and quality map}

For each journal in each window, performance is ranked on each of the seven indicators (1 = best within the FT50, 53 = worst). The composite rank is the mean of the within-FT50 ranks across the seven indicators, then re-ranked. Within-FT50 quartiles of the composite are used for tier interpretation. To complement the rank-based composite with a magnitude-aware visualization, each indicator is also z-scored across the 53 journals and averaged within channel (citation indicators in one composite, policy and patent indicators in another); each journal is then placed on a two-dimensional plane defined by the citation z-composite (x-axis) and the societal z-composite (y-axis). This visualization is referred to as the FT50 quality map throughout the paper.

\section{Results}

\subsection{Per-publication impact varies by orders of magnitude across the FT50}

Even after per-publication normalization, the journals on the list span very large ranges on every channel. In the 2005--2019 window, the \textit{Quarterly Journal of Economics} records 88.3 policy citations per published paper while \textit{Contemporary Accounting Research} records 0.36, a span of roughly 250-fold. On patent citations per paper, \textit{MIS Quarterly} records 0.116 and the \textit{Academy of Management Review} records essentially zero. On FWCI the span is narrower but still wide: \textit{Academy of Management Annals} at approximately 8.4, \textit{Operations Research} at approximately 1.16, a factor of seven. On the share of papers ever reaching policy, the elite economics journals lead (\textit{American Economic Review} at 90\%, \textit{Quarterly Journal of Economics} at 98\%); practitioner outlets and several accounting and management journals trail at single-digit percentages. These ranges hold across all three replication windows.

\begin{figure}[ht]
\centering
\includegraphics[width=\linewidth]{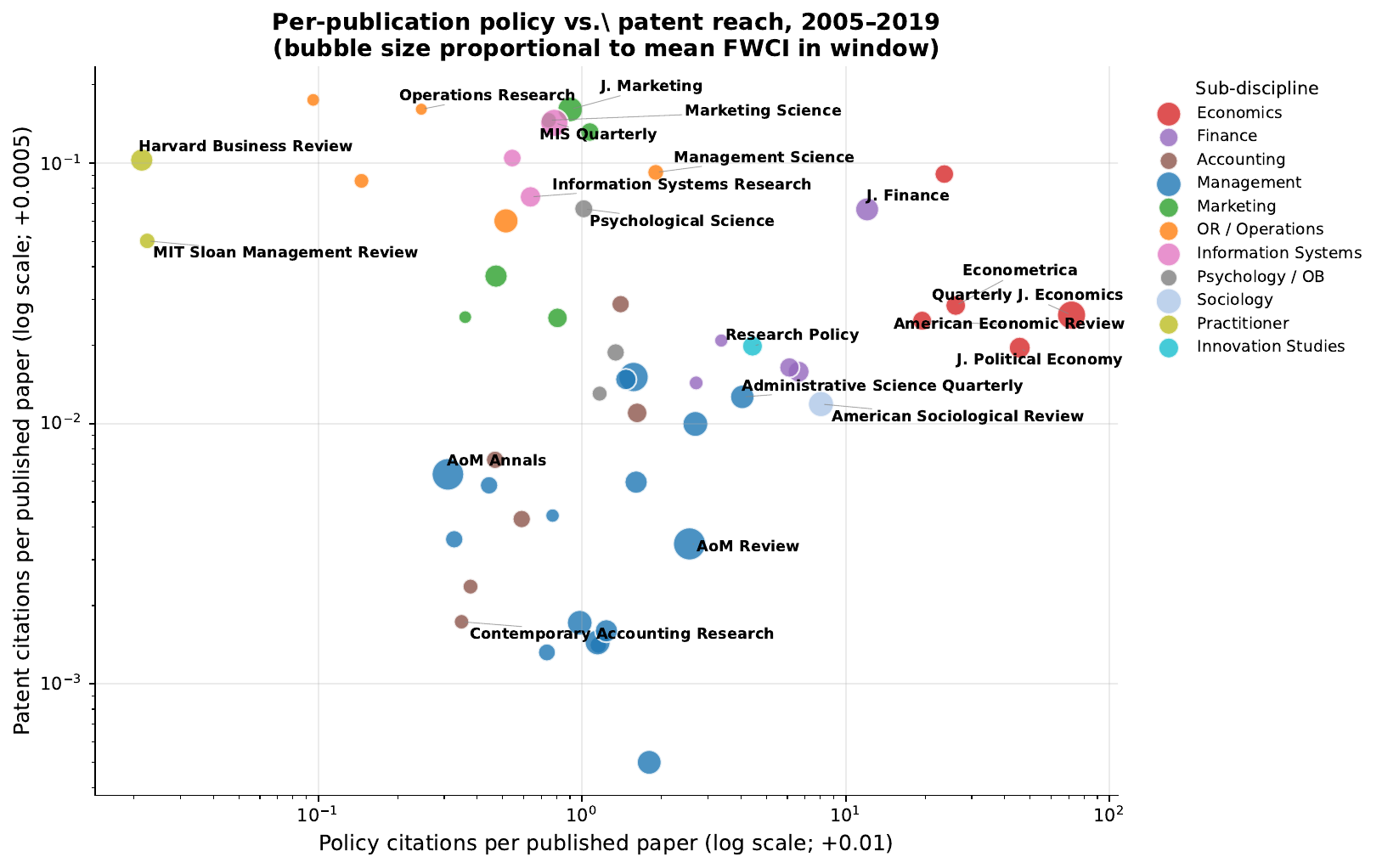}
\caption{Per-publication policy reach (x-axis, log) versus per-publication patent reach (y-axis, log) for the 53 FT50 journals in the 2005--2019 publication-year window. Bubble size is proportional to mean FWCI in the window; colour indicates sub-discipline. The three populated regions correspond to the policy channel (lower right), the patent channel (upper left), and a low-reach corner (lower left).}
\label{fig:per_pub}
\end{figure}

The map in Figure~\ref{fig:per_pub} plots each journal in (per-publication policy reach, per-publication patent reach) space, with bubble size proportional to mean FWCI in the window. The picture is a three-region structure. The lower-right quadrant is the policy channel, occupied by the elite economics journals (\textit{Quarterly Journal of Economics}, \textit{Journal of Political Economy}, \textit{American Economic Review}, \textit{Econometrica}), the \textit{Journal of Finance}, and a handful of journals that achieve genuine policy reach despite different home disciplines (\textit{American Sociological Review}, \textit{Research Policy}). The upper-left quadrant is the patent channel, occupied by the IS journals (\textit{MIS Quarterly}, \textit{Information Systems Research}), the marketing science journals (\textit{Marketing Science}, \textit{Journal of Marketing}, \textit{Journal of Marketing Research}), the practitioner outlets (\textit{Harvard Business Review}, \textit{MIT Sloan Management Review}), and a small set of operations research journals. The lower-left corner contains journals whose per-paper reach is low on both societal channels: a sizeable share of the management, accounting, and organizational-behaviour outlets. The upper-right corner is empty. No FT50 journal achieves both high per-paper policy reach and high per-paper patent reach.

\subsection{The FT50 quality map: where each journal sits on a single 2D plane}
\label{sec:qmap}

\begin{figure}[ht]
\centering
\includegraphics[width=\linewidth]{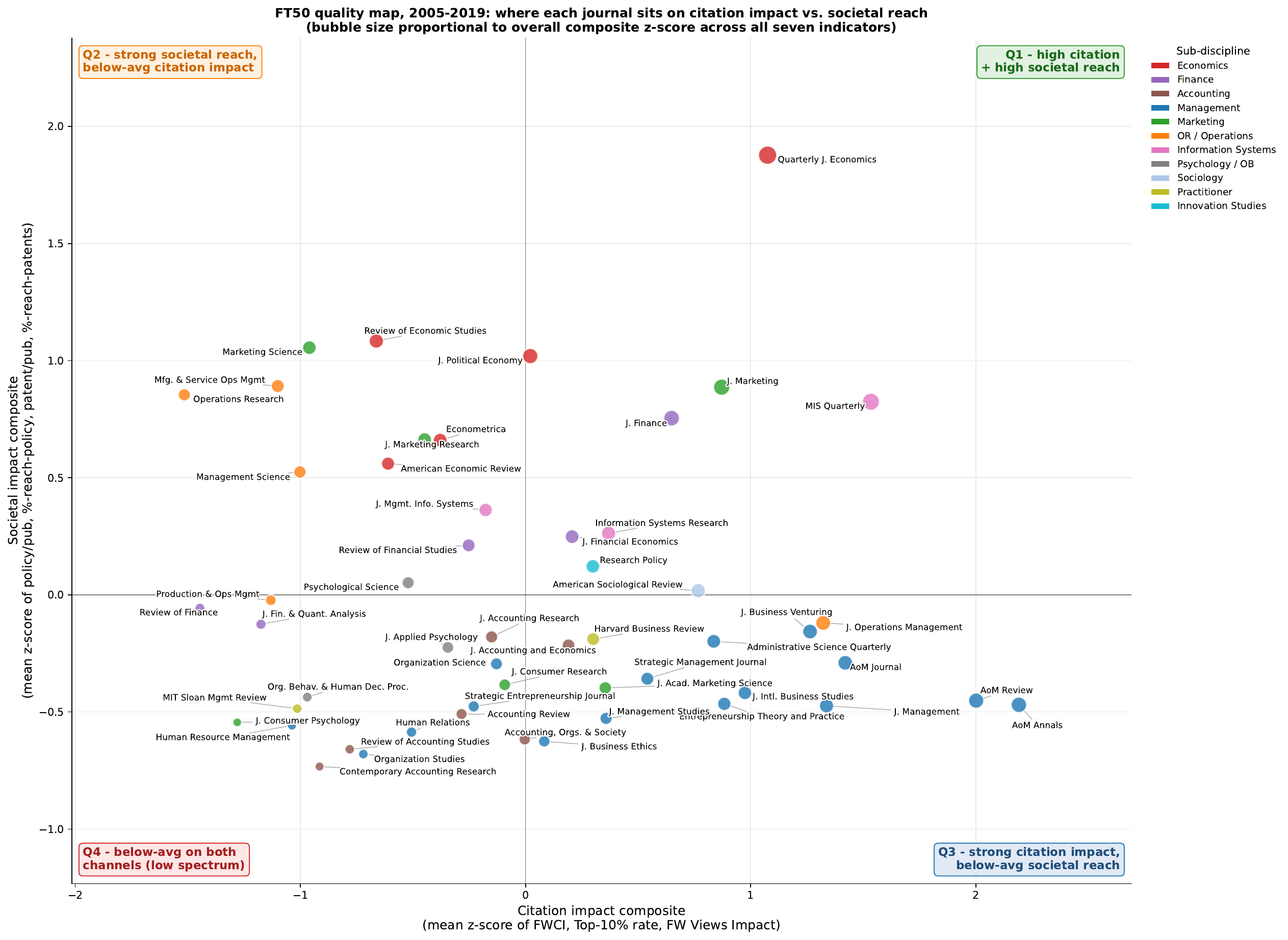}
\caption{FT50 quality map, 2005--2019. Horizontal axis: citation-impact composite (mean z-score of FWCI, Top-10\% rate, FW Views Impact). Vertical axis: societal-impact composite (mean z-score of policy citations per pub, share reaching policy, patent citations per pub, share reaching patents). Bubble size is proportional to the overall z-score across all seven indicators. Colour indicates sub-discipline. Quadrant lines fall at zero on each axis.}
\label{fig:qmap}
\end{figure}

Figure~\ref{fig:qmap} places every FT50 journal on a single quality map. The horizontal axis is the journal's average z-score on the three citation-impact indicators (FWCI, Top-10\% rate, Field-Weighted Views Impact). The vertical axis is the average z-score on the four societal-impact indicators (policy citations per pub, share reaching policy, patent citations per pub, share reaching patents). Both axes are standardized across the 53 FT50 journals, so a value of 0 is the FT50 average; positive means above average, negative means below average. Bubble size is proportional to the overall z-score across all seven indicators, so the larger the bubble the higher the journal's all-channel quality.

The four quadrants of the map carry distinct meanings. Q1 (upper right, n = 9) contains journals above the FT50 average on both citation impact and societal reach: the \textit{Quarterly Journal of Economics} is the standout (cit z = +1.07, soc z = +1.88), followed by \textit{MIS Quarterly}, \textit{Journal of Marketing}, \textit{Journal of Finance}, \textit{Journal of Political Economy}, \textit{American Sociological Review}, \textit{Information Systems Research}, \textit{Journal of Financial Economics}, and \textit{Research Policy}. These nine sit at the top of the spectrum across every channel. Q2 (upper left, n = 11) contains journals strong on societal reach but below the within-FT50 citation average. The \textit{American Economic Review}, \textit{Econometrica}, and \textit{Review of Economic Studies} all sit here, not because their citation impact is low in absolute terms but because the FT50 is a list of high-citation journals to begin with: a journal can rank in the world's top decile of business research and still fall below the FT50's internal average. Q3 (lower right, n = 15) is the citation-prestige cluster, with \textit{Academy of Management Annals} as the rightmost outlier (cit z = +2.19) and the rest of the cluster dominated by management journals with strong field-normalized citation impact and weak external reach. Q4 (lower left, n = 18) contains journals below the FT50 average on both channels.

\subsection{The quality spectrum: a stable identification of top and bottom tiers}
\label{sec:tiers}

\begin{figure}[p]
\centering
\includegraphics[width=\linewidth,height=0.88\textheight,keepaspectratio]{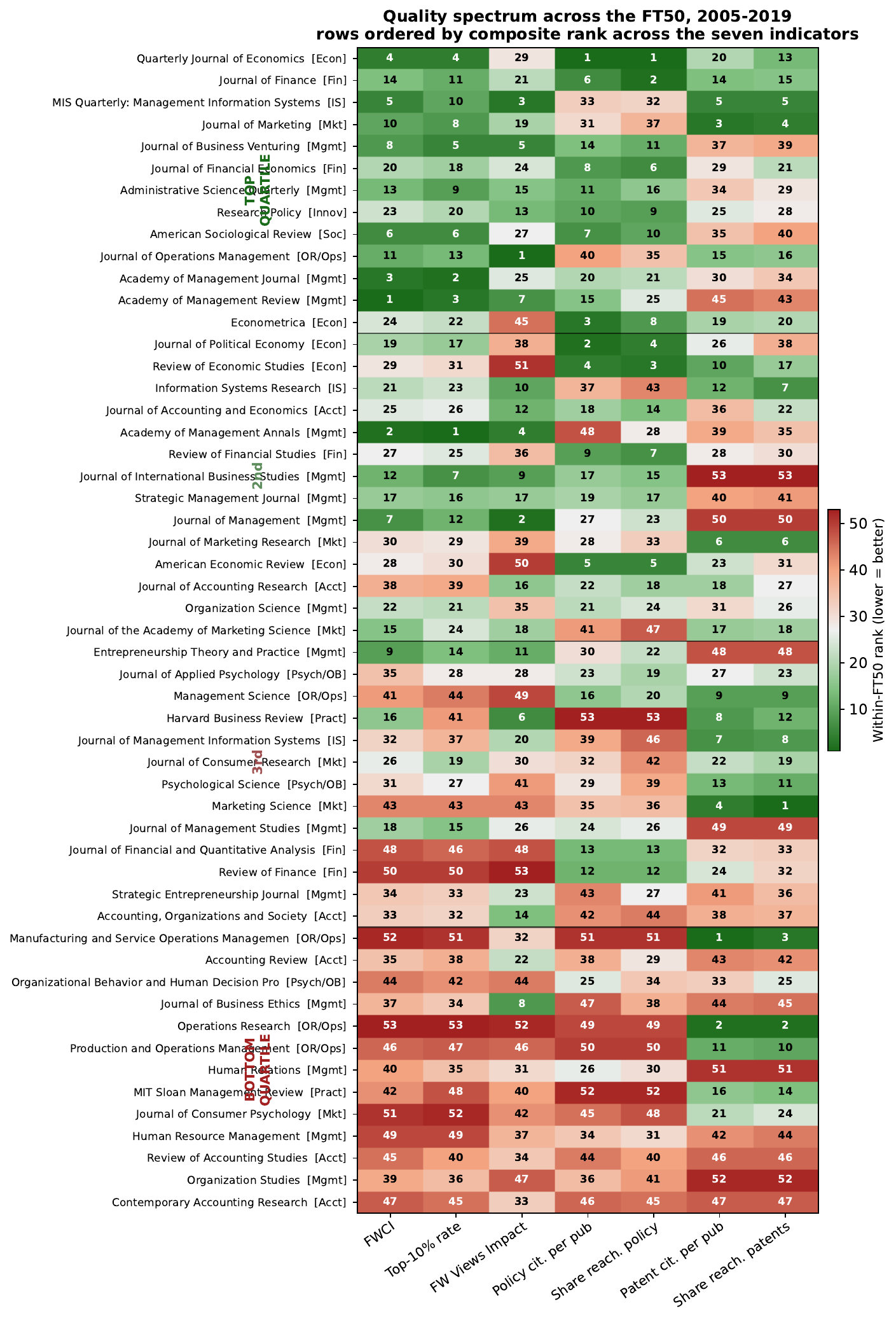}
\caption{Within-FT50 rank of each journal on each of the seven per-publication indicators in the 2005--2019 window. Rows are sorted by composite rank. Cell colour and number indicate the journal's rank from 1 (best, dark green) to 53 (worst, dark red). Horizontal lines mark the boundaries between within-FT50 quartiles.}
\label{fig:spectrum}
\end{figure}

Translating the map into a single ordering, each journal is ranked on each of the seven indicators and the mean rank is taken as a composite score (Section~\ref{sec:indicators}). The ordering, presented as a heatmap of within-FT50 ranks, is shown in Figure~\ref{fig:spectrum}.

The top quartile of the FT50 contains thirteen journals: \textit{Quarterly Journal of Economics} (composite rank 1), \textit{Journal of Finance}, \textit{MIS Quarterly}, \textit{Journal of Marketing}, \textit{Journal of Business Venturing}, \textit{Journal of Financial Economics}, \textit{Administrative Science Quarterly}, \textit{Research Policy}, \textit{American Sociological Review} and \textit{Journal of Operations Management} (tied at 9), \textit{Academy of Management Journal}, \textit{Academy of Management Review}, and \textit{Econometrica}. No journal in this group leads on every channel; what unites the thirteen is that each sits in the upper half of the list on at least four of the seven indicators. They span seven of the eleven sub-disciplines: economics, finance, information systems, management, marketing, sociology, and innovation studies. Two of the three 2026 additions (\textit{Academy of Management Annals} and \textit{American Sociological Review}) sit in the top half of the list, and the third (\textit{Psychological Science}) in the third quartile.

The bottom quartile contains thirteen journals: \textit{Manufacturing and Service Operations Management}, \textit{Accounting Review}, \textit{Organizational Behavior and Human Decision Processes}, \textit{Journal of Business Ethics}, \textit{Production and Operations Management}, \textit{Operations Research}, \textit{Human Relations}, \textit{MIT Sloan Management Review}, \textit{Journal of Consumer Psychology}, \textit{Human Resource Management}, \textit{Review of Accounting Studies}, \textit{Organization Studies}, and \textit{Contemporary Accounting Research}. Each sits in the lower half of the list on at least four of the seven indicators. The composition leans heavily on accounting (five of the six FT50 accounting journals are here), on operations (three of the five operations journals), and on the practitioner \textit{Sloan Management Review}. All three of the 2026-removed journals (\textit{Organization Studies}, \textit{Human Relations}, \textit{Journal of Business Ethics}) are in this bottom quartile.

\begin{figure}[ht]
\centering
\includegraphics[width=\linewidth]{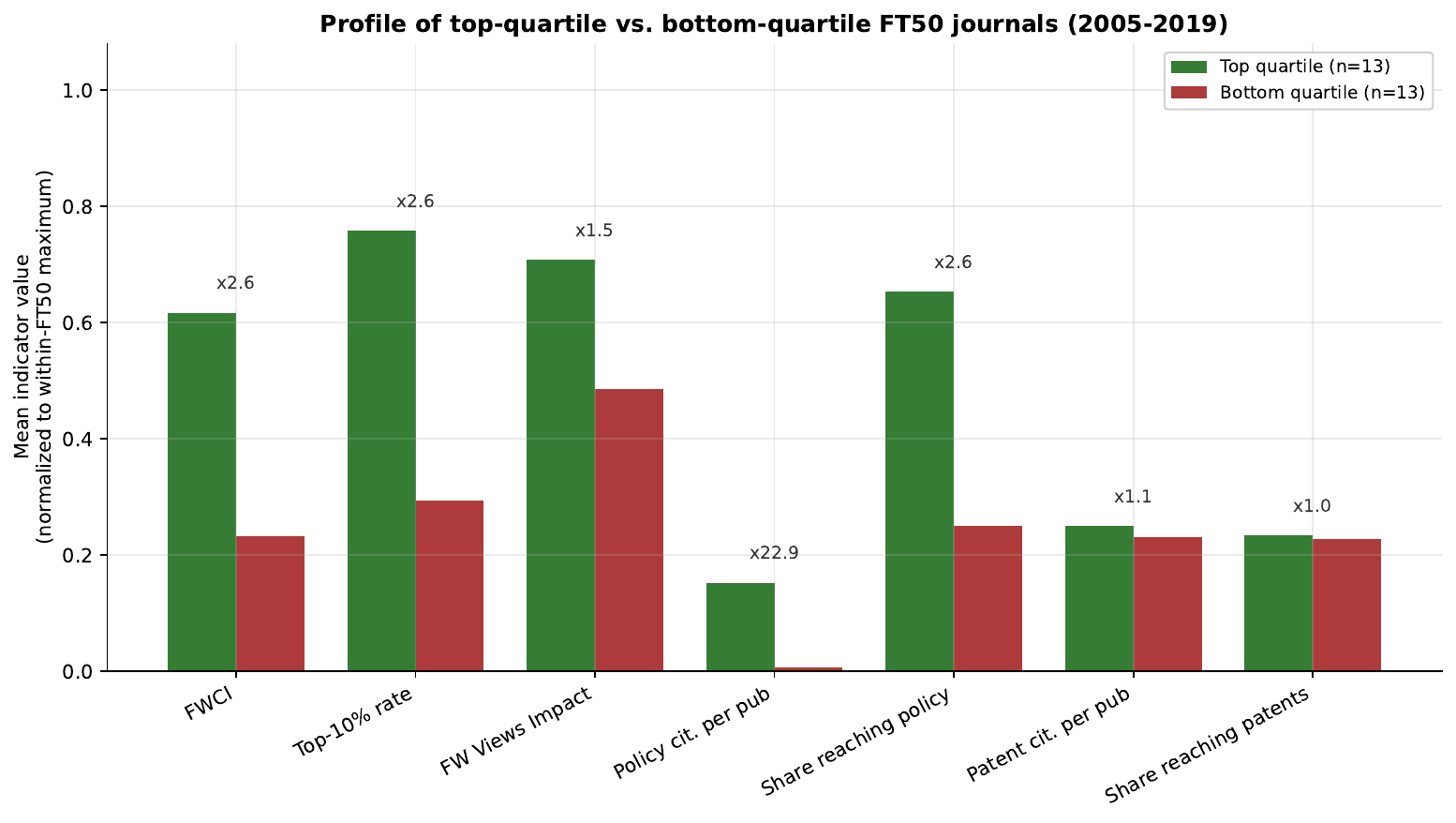}
\caption{Mean indicator value of the top and bottom within-FT50 quartiles, normalized to the within-FT50 maximum, with the multiplicative ratio between top and bottom annotated above each pair. The largest gap is on policy citations per published paper.}
\label{fig:topbot}
\end{figure}

The contrast between the top and bottom quartiles is large in absolute as well as ordinal terms (Figure~\ref{fig:topbot}). The multiplicative gap is largest on the policy channel: the top-quartile mean is roughly nineteen times the bottom-quartile mean on policy citations per pub. On FWCI, Top-10\% rate, and share-reaching-policy, the gap is roughly 2.5-fold. On patent reach the gap is smaller (between 1.5 and 2-fold), reflecting the broader cross-disciplinary distribution of industry uptake.

\subsection{The identification is stable across publication-year windows}
\label{sec:stability}

\begin{figure}[ht]
\centering
\includegraphics[width=\linewidth]{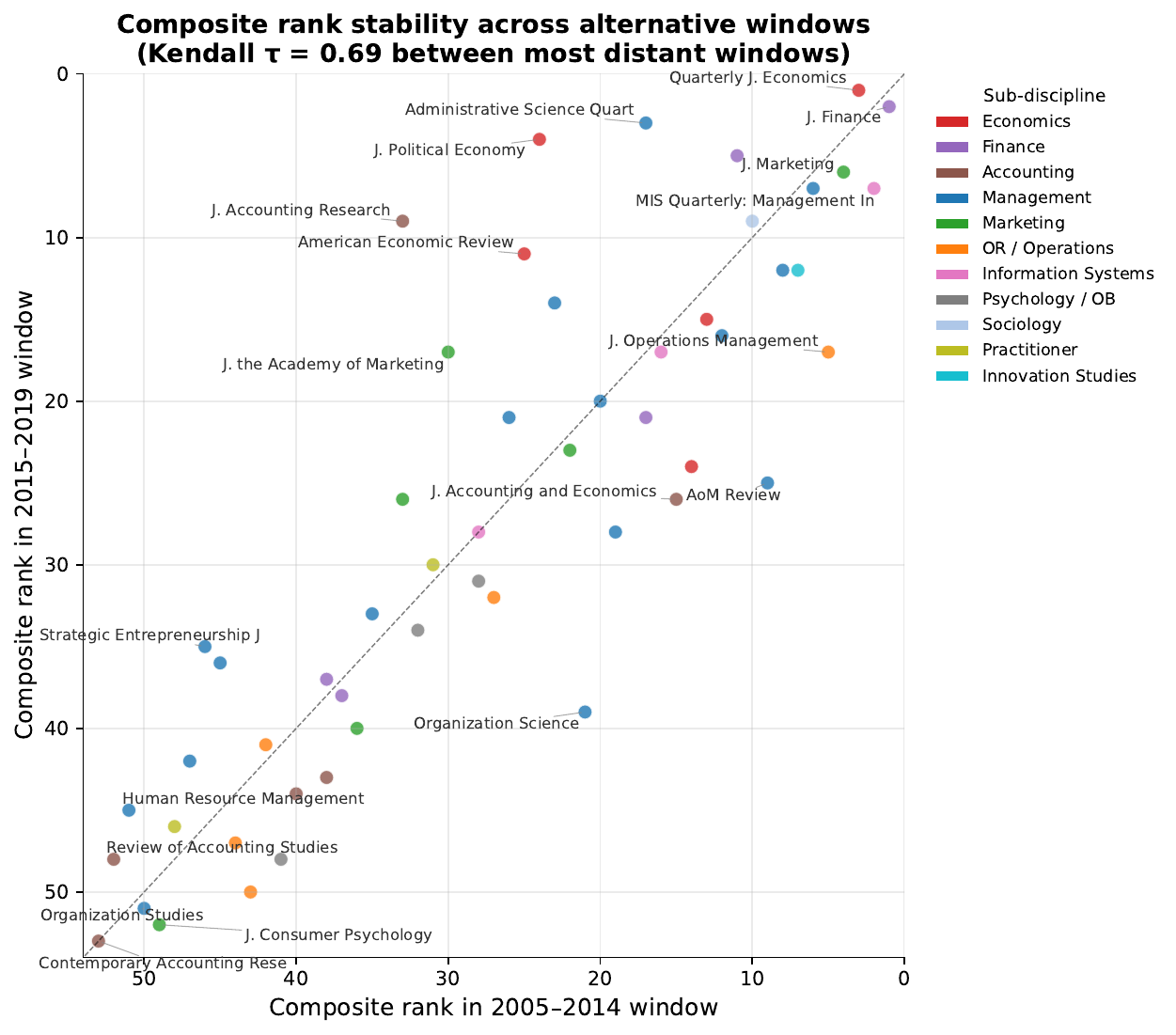}
\caption{Composite within-FT50 rank in the 2005--2014 window (x-axis) versus the 2015--2019 window (y-axis), with axes inverted so that rank 1 sits in the upper right. Points cluster tightly along the diagonal, indicating that the ranking is largely stable across the decade-long shift in window. A small number of journals (Journal of Accounting Research, Organization Science, Journal of Political Economy) move appreciably, suggesting genuine relative shifts in their indicator profiles.}
\label{fig:stability}
\end{figure}

The composite ranking might depend on the specific 2005--2019 window. To check this, the per-journal indicator computation is replicated under three alternative windows (2005--2014, 2010--2019, and 2015--2019) and journals are ranked on each indicator in each window separately. Figure~\ref{fig:stability} plots each journal's composite rank in the earliest and latest replication windows. Points cluster tightly along the diagonal: most journals occupy almost the same position in 2005--2014 as in 2015--2019.

\begin{figure}[ht]
\centering
\includegraphics[width=\linewidth]{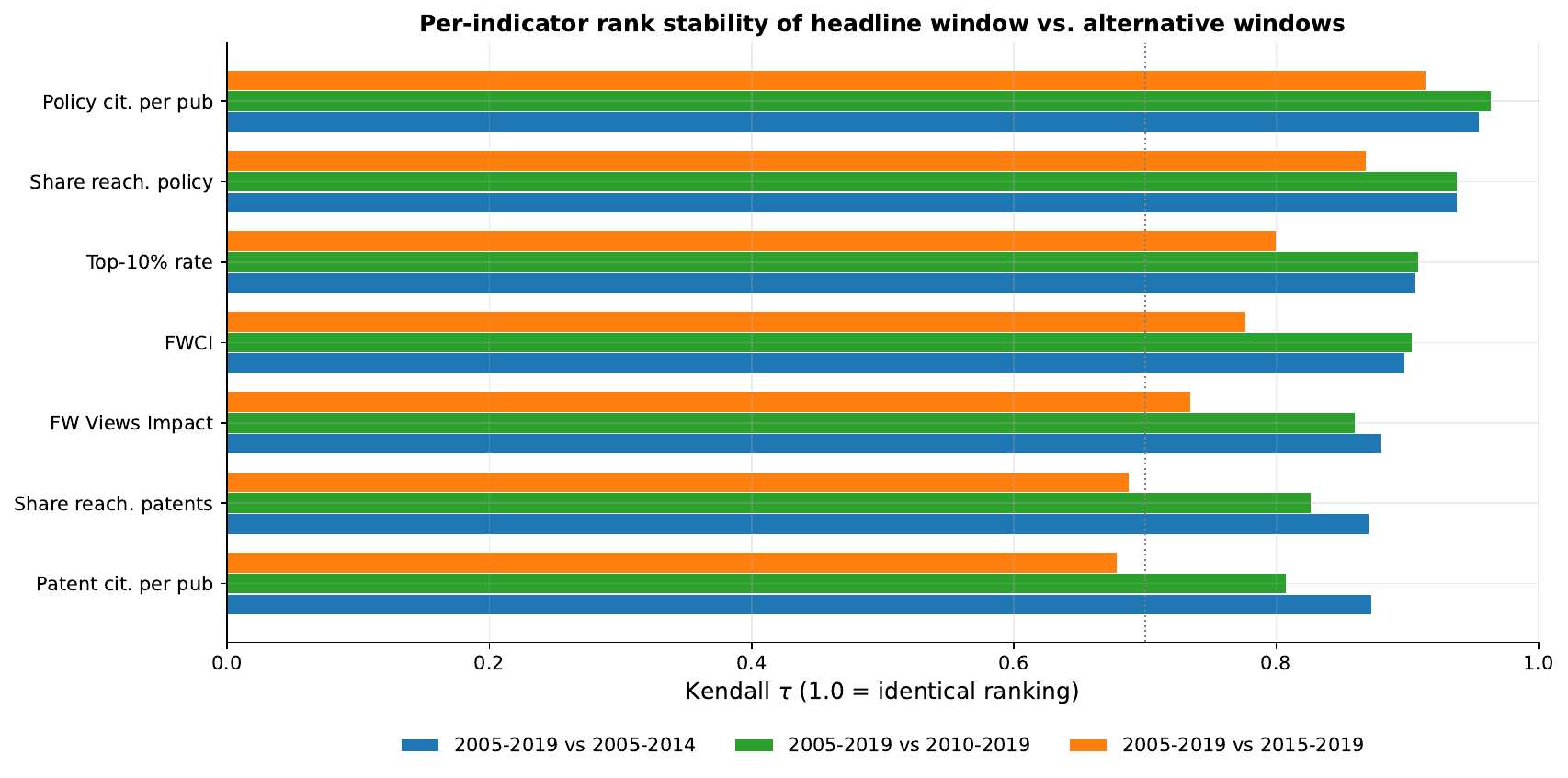}
\caption{Kendall $\tau$ between rankings produced under different publication-year windows, by indicator. The dotted vertical line marks $\tau = 0.7$. Policy-related indicators are the most stable; patent-related indicators are the least stable, reflecting their sparser distribution.}
\label{fig:sensitivity}
\end{figure}

The Kendall $\tau$ between the headline ranking and each of the three sensitivity windows is 0.89 (versus 2005--2014), 0.86 (versus 2010--2019), and 0.78 (versus 2015--2019). The $\tau$ between the most distant pair of sensitivity windows is 0.69 (2005--2014 versus 2015--2019). All four values are at levels read in the scientometrics literature as substantial agreement. Per-indicator stability (Figure~\ref{fig:sensitivity}) is highest for the policy measures, with $\tau$ above 0.91 against any sensitivity window. FWCI, the Top-10\% rate, and Field-Weighted Views Impact sit in the middle, with $\tau$ between 0.74 and 0.91. The patent indicators are the noisiest, with $\tau$ between 0.68 and 0.87, because patent reach is the sparsest signal in the panel. Within these bounds the tier assignment is stable: of the thirteen journals in the bottom quartile of the 2005--2019 ranking, eleven or more are also in the bottom quartile of every sensitivity window, and the corresponding numbers for the top quartile are at least eleven of thirteen.

The journals whose ranks move most across windows are themselves informative. \textit{Journal of Accounting Research} moves twelve ranks between the earliest and latest sensitivity windows; \textit{Journal of Political Economy} moves ten; \textit{Organization Science} moves nine. These movements do not weaken the comparison framework. They show that some journals are genuinely shifting their indicator profile over the past two decades, and a longitudinal extension that models within-journal change is a natural follow-up.

\subsection{Channel orthogonality holds at the per-publication level}

The cross-journal comparison demonstrates that citation, policy, and patent reach behave as nearly orthogonal channels across the FT50 on per-publication terms, consistent with the cross-channel decoupling discussed in the Overton literature \citep{Szomszor2022}. In the 2005--2019 window, the Spearman rank correlation across the 53 journals between FWCI and policy citations per pub is 0.27, between FWCI and patent citations per pub is $-0.12$, and between policy citations per pub and patent citations per pub is 0.08. The Top-10\% rate correlates more strongly with policy reach ($\rho = 0.63$) but only weakly with patent reach ($\rho = 0.24$). A journal's rank on one channel does not predict its rank on the others, even after per-publication adjustment. The empty upper-right quadrant of Figure~\ref{fig:per_pub} is the geometric expression of this point: no FT50 journal in the 2005--2019 window achieves above-median per-paper policy reach and above-median per-paper patent reach simultaneously.

The disciplinary composition of the channels matches the decoupling. Five economics journals account for 65\% of all FT50 policy-citation footprint over the window. Operations Research and Operations Management journals account for 33\% of the patent footprint; Marketing journals account for another 20\%. Management journals dominate raw citation count and views (32.7\% and 34.3\% respectively) but contribute less than 8\% to either of the societal channels.

\subsection{How distinct are the policy and patent measures from citation-based measures?}
\label{sec:distinct}

A reasonable objection to the seven-indicator composite is that the policy and patent measures may simply track the citation measures. If they do, the composite is a citation ranking in disguise, and the two societal channels add nothing. The question is whether the four societal indicators (policy citations per pub, share of papers reaching policy, patent citations per pub, share of papers reaching patents) carry information that the three citation indicators (FWCI, Top-10\% rate, Field-Weighted Views Impact) do not. The data answer it directly.

\begin{figure}[ht]
\centering
\includegraphics[width=\linewidth]{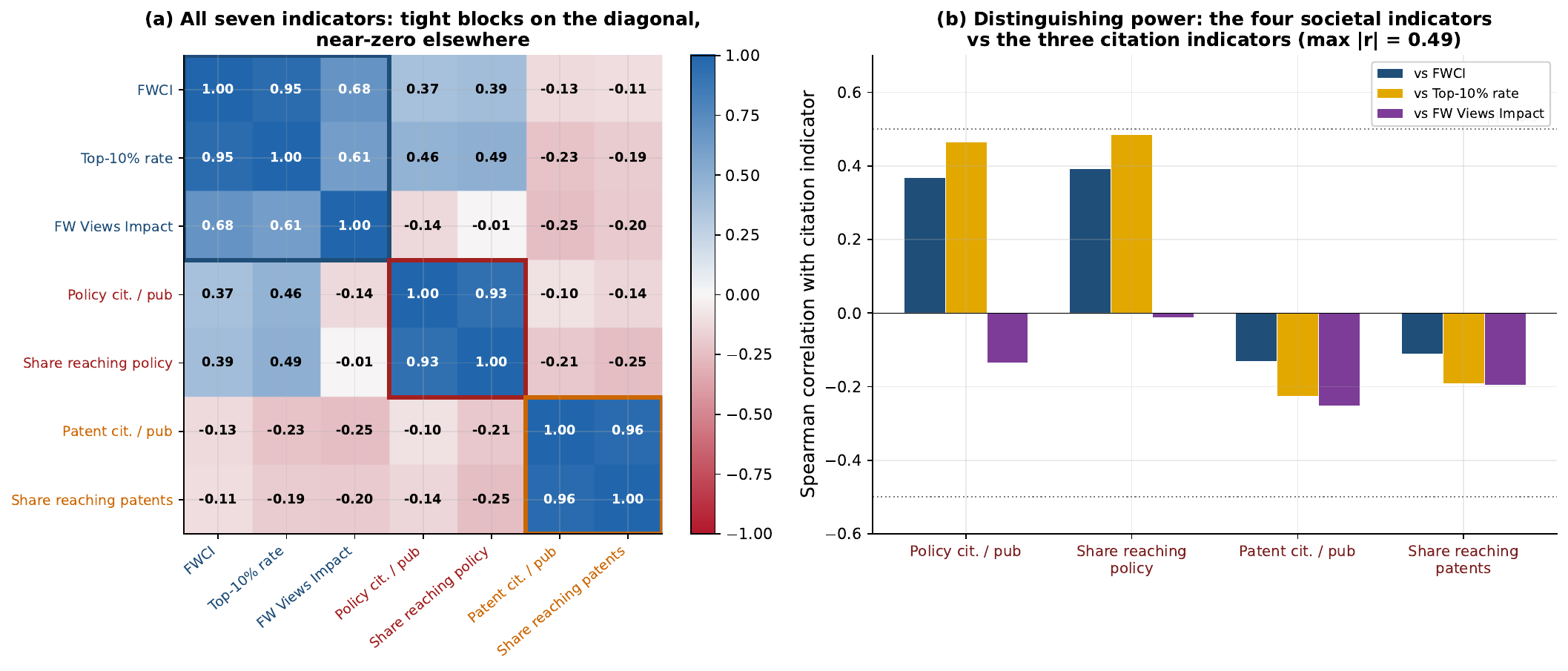}
\caption{Distinguishing power of the societal indicators. (a) Spearman correlation among all seven per-publication indicators, with channel blocks outlined; within-channel correlations average 0.83, across-channel correlations average $-0.02$. (b) Each of the four societal indicators against each of the three citation indicators; no pair exceeds $|r| = 0.49$, and the patent measures are weakly negative with all three citation measures.}
\label{fig:distinct}
\end{figure}

The strongest correlation between any societal indicator and any citation indicator is 0.49, between the share of papers reaching policy and the Top-10\% rate (Figure~\ref{fig:distinct}b). That is the maximum; it corresponds to 24\% shared variance, so even the most citation-aligned societal measure leaves three-quarters of its variance unexplained by the closest citation measure. The mean absolute correlation between a societal indicator and a citation indicator is 0.25. The two policy measures are weakly-to-moderately positive with FWCI and the Top-10\% rate (0.37 to 0.49) and near zero with Field-Weighted Views Impact; the two patent measures are weakly negative with all three citation measures ($-0.11$ to $-0.25$). On the patent side, a journal that scores well on citations tends, if anything, to score slightly worse on patent reach. Policy and patent reach are not citations under another name.

The full picture is shown in Figure~\ref{fig:distinct}a. The seven-indicator correlation matrix has a clear block structure: indicators within a channel are strongly collinear, which is the intended behaviour, because each pair measures one construct from two angles. The three citation indicators carry a mean pairwise correlation of 0.83, the two policy indicators correlate at 0.93, and the two patent indicators at 0.96. Across channels the mean correlation is $-0.02$. A principal component analysis confirms the dimensionality: three components are needed to account for 91\% of the variance in the seven indicators (43\%, 26\%, and 22\% respectively), and the loadings separate the channels rather than blending them. A single ``quality'' factor does not exist in these data; the FT50 occupies a three-dimensional impact space.

\begin{figure}[ht]
\centering
\includegraphics[width=0.92\linewidth]{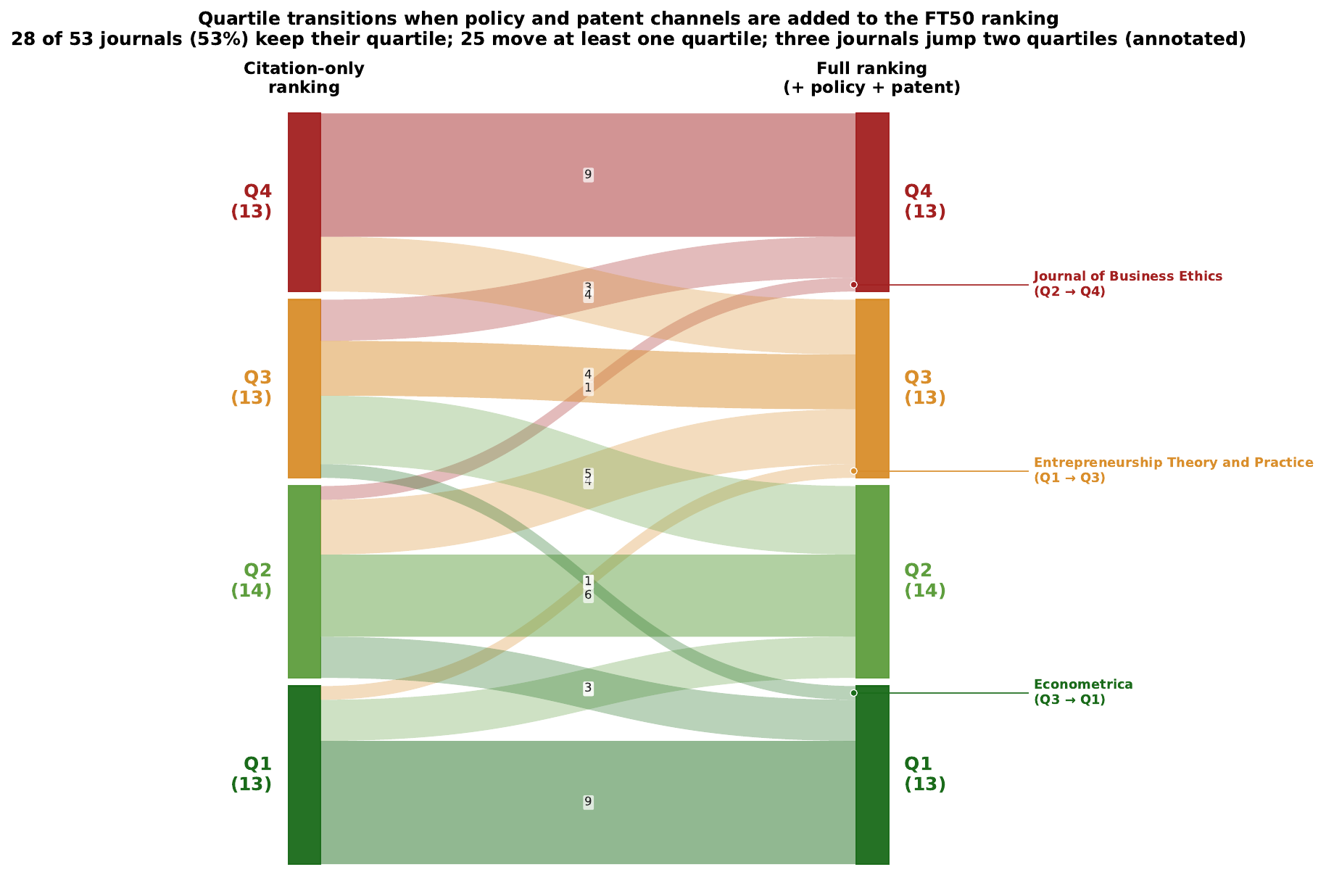}
\caption{Quartile transitions of the 53 FT50 journals between the citation-only ranking (left) and the full seven-indicator ranking (right). Ribbon width is proportional to the number of journals making each transition; the number on each ribbon is that count. Twenty-eight journals keep their quartile; twenty-five move at least one quartile.}
\label{fig:shift}
\end{figure}

The practical consequence is that the choice of indicator set changes the ranking. A citation-only composite, built from the three citation indicators alone, correlates with the full seven-indicator composite at a Kendall $\tau$ of only 0.54. Sorting the 53 journals into within-FT50 quartiles under each indicator set, 25 of the 53 move at least one quartile when the policy and patent channels are added, and only 28 stay put (Figure~\ref{fig:shift}). The largest single-journal moves run in both directions: the elite economics and finance journals rise (\textit{Review of Economic Studies} by 24 rank positions, \textit{Econometrica} by 18, \textit{Management Science} by 16, \textit{Review of Finance} by 15), while management journals whose strength is concentrated in the citation channel fall (\textit{Entrepreneurship Theory and Practice} by 19, \textit{Journal of Business Ethics} by 18, \textit{Academy of Management Annals} by 17, \textit{Journal of Management} by 17). The citation channel and the societal channels reward different journals. A ranking that includes only citations has made an implicit and consequential choice.

\subsection{Trajectories and near-term outlook}
\label{sec:trajectory}

The cross-sectional ranking describes where journals sit over 2005--2019 as a whole. A separate question is where they are heading. To address it, the seven-indicator composite ranking is recomputed on seventeen overlapping five-year windows, from 2005--2009 through 2021--2025, stepped one year at a time. Because each window is ranked internally, the common citation-accrual shift that affects all journals in a given window cancels out, and the resulting series of composite ranks is comparable across windows. Two journals founded mid-window, the \textit{Academy of Management Annals} and the \textit{Strategic Entrepreneurship Journal}, do not have a complete seventeen-window series and are excluded from the trajectory analysis, leaving 51 journals. The windows that include 2020 onward carry more accrual noise, particularly on the patent indicators, where citations to papers published after about 2022 are not yet observable; the within-window ranking absorbs the common component, and the composite is carried mainly by the citation and policy channels in those windows.

The trend line for each journal is computed by ordinary least squares: the journal's composite rank in each of the seventeen windows is regressed on the window's midpoint year. The fitted slope is the journal's trajectory, in rank positions per year; a negative slope means the journal is improving, because a lower rank number is better. The fitted line extended three years past the last window gives the projection. With the last window centred on 2023, the projection reaches a window midpoint of 2026, that is, a notional 2024--2028 window. It is a naive linear extrapolation, offered as an indication of direction rather than a forecast.

\begin{figure}[ht]
\centering
\includegraphics[width=\linewidth]{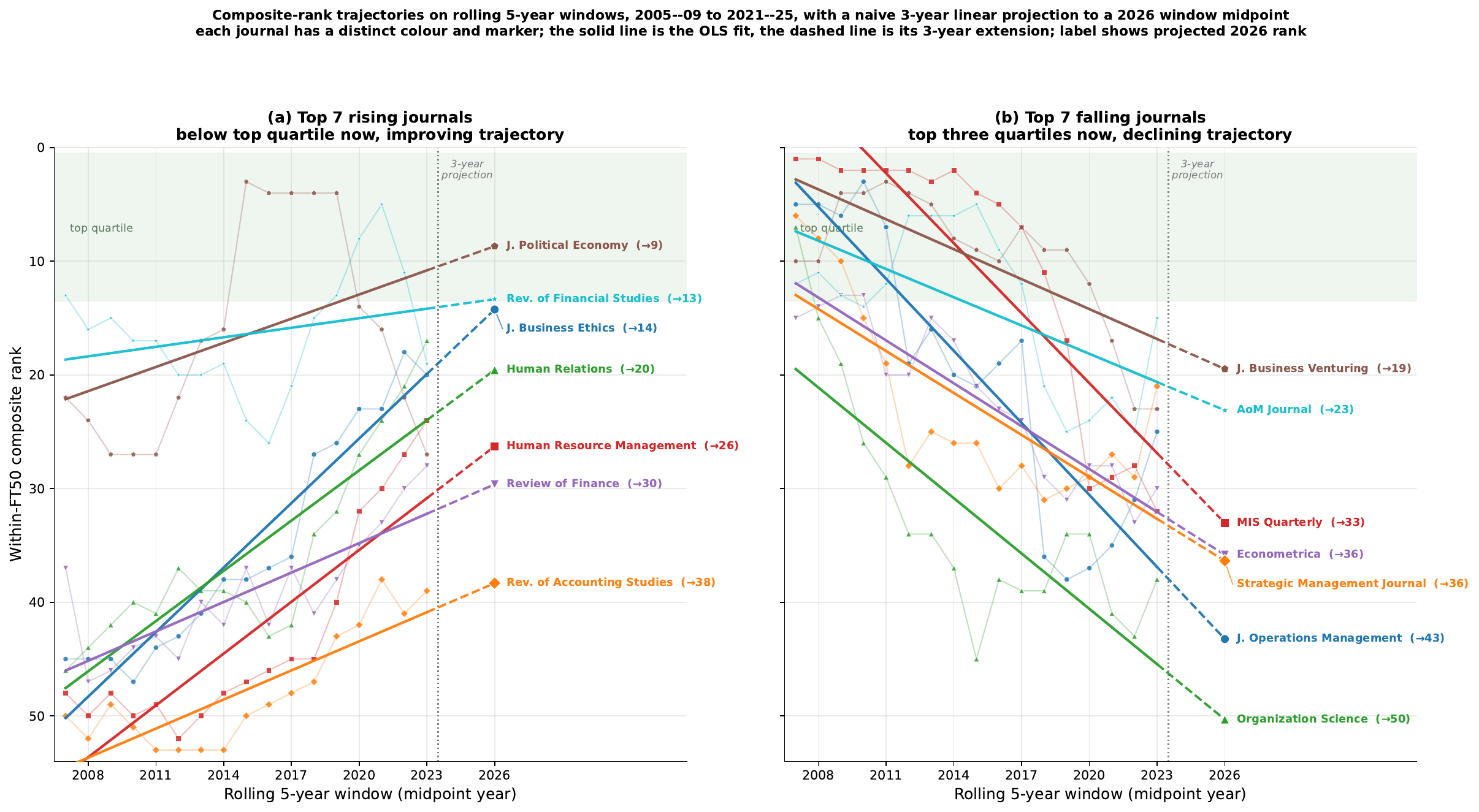}
\caption{Composite-rank trajectories on rolling five-year windows, 2005--2009 to 2021--2025, with a naive three-year linear projection (dashed) reaching a 2026 window midpoint. The trend line for each journal is an ordinary-least-squares fit of composite rank on window midpoint year. (a) The seven journals with the steepest improving trajectory that are not in the top quartile; their lines move up toward rank 1. (b) The seven journals with the steepest declining trajectory that are currently in the top three quartiles; their lines move down. Each label shows the journal and its projected 2026 rank in brackets, with a leader line to its trend-line endpoint. The shaded band marks the within-FT50 top quartile. Windows that include 2020--2025 carry more accrual noise, especially on the patent channel.}
\label{fig:trajectory}
\end{figure}

Two groups are of interest. The first is journals not currently in the top quartile but with an improving trajectory (negative slope); the seven steepest are the \textit{Journal of Business Ethics} ($-1.89$ rank positions per year), \textit{Human Resource Management} ($-1.52$), \textit{Human Relations} ($-1.47$), the \textit{Review of Finance} ($-0.86$), the \textit{Review of Accounting Studies} ($-0.85$), the \textit{Journal of Political Economy} ($-0.71$), and the \textit{Review of Financial Studies} ($-0.28$). The \textit{Journal of Political Economy}, currently at rank 27, has the most consequential improving trajectory, projecting toward the top quartile by 2026. The second group is journals currently in the top three quartiles but with a declining trajectory (positive slope); the seven steepest are the \textit{Journal of Operations Management} ($+2.11$ per year), \textit{MIS Quarterly} ($+2.05$), \textit{Organization Science} ($+1.62$), \textit{Econometrica} ($+1.26$), the \textit{Strategic Management Journal} ($+1.23$), the \textit{Journal of Business Venturing} ($+0.88$), and the \textit{Academy of Management Journal} ($+0.83$). Figure~\ref{fig:trajectory} plots both groups, with the linear projection extended three years past the last window.

The trajectory result is distinct from the ranking-shift result of the previous subsection and should not be conflated with it. The ranking shift is cross-sectional: it asks which journals gain when the indicator set is broadened. The trajectory is temporal: it asks which journals are moving up or down over time on the fixed seven-indicator composite. \textit{Econometrica}, for instance, gains 18 rank positions when the societal channels are added to the indicator set, yet its rolling-window trajectory is declining. Both statements describe the same journal without contradiction. The projections in Figure~\ref{fig:trajectory} assume that the recent linear trend continues, which it need not; they are best read as a structured way of flagging journals whose recent direction is worth watching, not as a model of future performance.

\section{Discussion: the 2026 list update against the comparative quality spectrum}

The Financial Times announced a revision of the list in 2026, the first substantial change since the 2016 reshuffling discussed in detail by \citet{Fassin2021}. Three journals were removed (\textit{Human Relations}, \textit{Journal of Business Ethics}, and \textit{Organization Studies}, all in the Management cluster) and three were added (\textit{Academy of Management Annals} in Management, \textit{American Sociological Review} in Sociology, and \textit{Psychological Science} in Psychology). All three of the removed journals sit in the bottom quartile of the 2005--2019 composite ranking: \textit{Organization Studies} at rank 52 of 53, \textit{Human Relations} at rank 47, and \textit{Journal of Business Ethics} at rank 44. On the quality map, \textit{Organization Studies} and \textit{Human Relations} sit in the lower-left Q4. \textit{Journal of Business Ethics} sits just inside Q3 with citation impact at the FT50 average and societal reach below it. Two of the three additions sit in the top half of the ranking: \textit{American Sociological Review} enters in the top quartile and the Q1 high-citation high-societal quadrant, and \textit{Academy of Management Annals} enters in the second quartile with an exceptional within-FT50 citation profile and below-average societal reach (Q3). \textit{Psychological Science} enters in the third quartile, with below-average citation impact and slightly above-average societal reach (Q2).

The swap tracks the comparative quality picture closely. The peer-perception judgement implicit in the FT's dean polls and editorial review is converging with the bibliometric picture the framework recovers, and the convergence is itself a useful empirical finding. The form of the swap, however, mostly tightens the bottom of the list and broadens the top. The citation channel strengthens through the addition of \textit{Annals}, a top-FWCI journal. The policy channel strengthens marginally through the addition of \textit{American Sociological Review}. The patent channel picks up a modest non-OR contribution through \textit{Psychological Science}. None of the three additions occupies the empty upper-right region of the quality map (Figure~\ref{fig:qmap}), where a journal would reach both policy and industry at scale at the same time. If the policy and patent dimensions of impact are part of what FT50 inclusion is meant to certify, the 2026 swap refines the existing impact profile of the list rather than broadening it.

A more fundamental implication concerns the structure of the list. The analysis identifies a within-FT50 quality spectrum that the binary in-or-out structure obscures. A channel-aware or multi-tier presentation of the list would communicate this heterogeneity to deans, accreditation bodies, and hiring committees without changing the list's role in research evaluation. The 2026 update is consistent with this reading: the FT appears to be making implicit channel-aware judgements when it removes bottom-quartile management journals and adds an exceptional citation-prestige journal alongside one policy-strong and one patent-strong addition.

\section{Caveats}
\label{sec:caveats}

Five caveats apply. First, the analysis is within-FT50. Without a control set of comparable elite non-FT50 journals (the natural candidate is the ABS-AJG 4* set) it is not possible to say whether the FT50 itself is moving relative to the wider business-research journal population. The aggregate trends observed (median FWCI of 3.45 in 2005 falling to 2.04 in 2020; total annual output growing 51\% over the same period) fit the citation-metric ecology that \citet{Hanson2024} and \citet{Fire2019} describe at the level of all of academic publishing, and are not necessarily features of the FT50 specifically.

Second, the SciVal extracts include self-citations by default. Self-citations have not been stripped from any of the indicators, and a self-citation-stripped variant would shift the rankings, particularly in fields with unusually high self-citation rates \citep{Lariviere2019,Ibrahim2025}. This is the cleanest methodological extension of the present analysis.

Third, the policy-citation infrastructure is itself growing. Overton has been adding coverage and depth across the windows used in the analysis \citep{Szomszor2022,Cristofoletti2025}, and a small share of the cross-journal heterogeneity in policy reach probably reflects uneven indexing across policy producers and source languages. The headline window starts in 2005 to mitigate the issue, but it is not eliminated.

Fourth, FT50 list membership is itself a moving target. The panel uses the current list, including the 2026 additions; all six swap journals are present because the SciVal extraction predates the FT's operationalization of the change. The longitudinal stability claim concerns the trajectory of journals that are or were recently on the list, not the trajectory of FT50-membership-as-a-set.

Fifth, the sub-discipline grouping is heuristic at the boundaries. The channel-share results are stable under alternative groupings (placing \textit{Journal of International Business Studies} in Economics rather than Management, and placing \textit{Research Policy} in Management rather than Innovation Studies). The full mapping is in Appendix~\ref{app:disc}.

\section{Conclusion}

When the FT50 is examined on per-publication, properly normalized terms, the heterogeneity it bundles is directly visible. The leading economics journals attract policy citations at a per-paper rate two orders of magnitude greater than the lowest journals on the list. The leading information-systems, marketing, and operations journals attract industry citations at a per-paper rate two to three orders of magnitude greater than the academic-management or accounting cluster. The composite ranking built from seven per-publication indicators identifies a top quartile of thirteen journals and a bottom quartile of thirteen journals whose membership is stable across alternative publication-year windows, with Kendall $\tau$ between rankings in the range 0.78 to 0.89 and the worst pair of sensitivity windows still at 0.69. The three impact channels are nearly orthogonal across the list: a journal's strength on one does not predict its strength on the others, and no journal achieves above-average per-paper policy reach and above-average per-paper patent reach simultaneously.

FT50 membership is not meaningless. The elite economics, finance, information-systems, and a small number of management and marketing journals do outperform across multiple channels. The substantive implication is that the in-or-out structure conceals a quality spectrum, and that a channel-aware presentation of the list (a small number of tiers based on which channel a journal leads) would communicate this heterogeneity directly. The 2026 update is consistent with this reading: it tightens the composite ranking at the top and the bottom while leaving the channel orthogonality of the list undisturbed.

\section*{Author contributions (CRediT)}

\noindent\textbf{Arash Hajikhani:} Conceptualization, Data curation, Formal analysis, Funding acquisition, Investigation, Methodology, Project administration, Resources, Software, Supervision, Validation, Visualization, Writing -- original draft, Writing -- review \& editing.
\textbf{Yi Zhang:} Investigation, Methodology, Resources, Software, Supervision, Validation, Writing -- original draft, Writing -- review \& editing.
\textbf{Mengjia Wu:} Investigation, Methodology, Validation, Visualization, Writing -- review \& editing.

\section*{Data availability statement}

The reshaped journal-year-metric panel (\texttt{ft50\_panel.csv}, 39{,}432 rows), the per-window per-journal indicator tables (2005--2019 headline, 2005--2014, 2010--2019, and 2015--2019 sensitivity replications), the composite-ranking and Kendall-$\tau$ stability tables, and all code used for the analysis and figures will be archived at Zenodo with a DOI on acceptance. The raw SciVal CSV exports cannot be shared openly because they are derived from a proprietary database (Scopus and SciVal); the indicator-extraction parameters and the 53-journal Scopus source-id list documented in the supplementary code allow exact replication for any user with SciVal access. Overton-derived policy citations are similarly proprietary in their full granularity; the aggregated per-journal indicators are computed and shared in the per-window indicator tables.

\bibliography{references}

@article{Baas2020,
  author  = {Baas, Jeroen and Schotten, Michiel and Plume, Andrew and C{\^o}t{\'e}, Gr{\'e}goire and Karimi, Reza},
  title   = {Scopus as a curated, high-quality bibliometric data source for academic research in quantitative science studies},
  journal = {Quantitative Science Studies},
  year    = {2020},
  volume  = {1},
  number  = {1},
  pages   = {377--386},
  doi     = {10.1162/qss_a_00019}
}

@article{Visser2021,
  author  = {Visser, Martijn and van Eck, Nees Jan and Waltman, Ludo},
  title   = {Large-scale comparison of bibliographic data sources: {S}copus, {W}eb of {S}cience, {D}imensions, {C}rossref, and {M}icrosoft {A}cademic},
  journal = {Quantitative Science Studies},
  year    = {2021},
  volume  = {2},
  number  = {1},
  pages   = {20--41},
  doi     = {10.1162/qss_a_00112}
}

@article{Szomszor2022,
  author  = {Szomszor, Martin and Adie, Euan},
  title   = {{Overton}: A bibliometric database of policy document citations},
  journal = {Quantitative Science Studies},
  year    = {2022},
  volume  = {3},
  number  = {3},
  pages   = {624--650},
  doi     = {10.1162/qss_a_00204}
}

@article{Hanson2024,
  author  = {Hanson, Mark A. and G{\'o}mez Barreiro, Pablo and Crosetto, Paolo and Brockington, Dan},
  title   = {The strain on scientific publishing},
  journal = {Quantitative Science Studies},
  year    = {2024},
  volume  = {5},
  number  = {4},
  pages   = {823--843},
  doi     = {10.1162/qss_a_00327}
}

@article{Cristofoletti2025,
  author  = {Cristofoletti, Evandro Coggo and Salles-Filho, Sergio and Juk, Yohanna and Cabral, Bernardo and Pinto, Karen Esteves Fernandes and Hollanda, Sandra and Graziani, Carlos and Pereira, C{\'e}sar Antonio},
  title   = {A long and winding road: Research impact evaluation over public policies},
  journal = {Quantitative Science Studies},
  year    = {2025},
  volume  = {6},
  pages   = {194--215},
  doi     = {10.1162/qss_a_00345}
}

@article{Mingers2017,
  author  = {Mingers, John and Yang, Liying},
  title   = {Evaluating journal quality: A review of journal citation indicators and ranking in business and management},
  journal = {European Journal of Operational Research},
  year    = {2017},
  volume  = {257},
  number  = {1},
  pages   = {323--337},
  doi     = {10.1016/j.ejor.2016.07.058}
}

@article{Vidgen2019,
  author  = {Vidgen, Richard and Mortenson, Michael and Powell, Philip},
  title   = {Invited Viewpoint: How well does the Information Systems discipline fare in the {Financial Times}' top 50 Journal list?},
  journal = {Journal of Strategic Information Systems},
  year    = {2019},
  volume  = {28},
  number  = {4},
  pages   = {101577},
  doi     = {10.1016/j.jsis.2019.101577}
}

@incollection{Lariviere2019,
  author    = {Larivi{\`e}re, Vincent and Sugimoto, Cassidy R.},
  title     = {The Journal Impact Factor: A brief history, critique, and discussion of adverse effects},
  booktitle = {Springer Handbook of Science and Technology Indicators},
  editor    = {Gl{\"a}nzel, Wolfgang and Moed, Henk F. and Schmoch, Ulrich and Thelwall, Mike},
  publisher = {Springer},
  year      = {2019},
  pages     = {3--24},
  doi       = {10.1007/978-3-030-02511-3_1}
}

@article{Meho2025,
  author  = {Meho, Lokman I.},
  title   = {Gaming the Metrics? Bibliometric Anomalies and the Integrity Crisis in Global University Rankings},
  journal = {bioRxiv preprint},
  year    = {2025},
  doi     = {10.1101/2025.05.09.653229}
}

@article{Fire2019,
  author  = {Fire, Michael and Guestrin, Carlos},
  title   = {Over-Optimization of Academic Publishing Metrics: Observing {G}oodhart's Law in Action},
  journal = {GigaScience},
  year    = {2019},
  volume  = {8},
  number  = {6},
  pages   = {giz053},
  doi     = {10.1093/gigascience/giz053}
}

@misc{JackDalal2024,
  author       = {Jack, Andrew and Dalal, Anjali},
  title        = {Business school and the pursuit of rigour, resonance and relevance},
  howpublished = {Financial Times},
  year         = {2024},
  month        = nov,
  note         = {Published 2024-11-08}
}

@article{Ibrahim2025,
  author  = {Ibrahim, Hazem and Liu, Fengyuan and Zaki, Yasir and Rahwan, Talal},
  title   = {Citation manipulation through citation mills and pre-print servers},
  journal = {Scientific Reports},
  year    = {2025},
  volume  = {15},
  pages   = {5480},
  doi     = {10.1038/s41598-025-88709-7}
}

@article{Zhang2021,
  author  = {Zhang, Tenghao},
  title   = {Will the increase in publication volumes ``dilute'' prestigious journals' impact factors? {A} trend analysis of the {FT50} journals},
  journal = {Scientometrics},
  year    = {2021},
  volume  = {126},
  pages   = {863--869},
  doi     = {10.1007/s11192-020-03736-7}
}

@article{Fassin2021,
  author  = {Fassin, Yves},
  title   = {Does the {Financial Times FT50} journal list select the best management and economics journals?},
  journal = {Scientometrics},
  year    = {2021},
  volume  = {126},
  pages   = {5911--5943},
  doi     = {10.1007/s11192-021-03988-x}
}

@article{Steingard2026,
  author  = {Steingard, David and Linacre, Simon and Reibstein, David and Flynn, Dylan and Springer, Caden},
  title   = {Revolutionizing societal impact in business school research: can the {FT50} lead the change?},
  journal = {Society and Business Review},
  year    = {2026},
  volume  = {21},
  number  = {2},
  pages   = {316--333},
  doi     = {10.1108/SBR-08-2025-0334}
}

@misc{ElsevierPatent2019,
  author       = {{Elsevier B.V.}},
  title        = {{SciVal} Usage and Patent Metrics Guidebook},
  howpublished = {Research Intelligence, Elsevier},
  year         = {2019}
}

@misc{ElsevierMetrics2019,
  author       = {{Elsevier B.V.}},
  title        = {Research Metrics Guidebook},
  howpublished = {Research Intelligence, Elsevier},
  year         = {2019}
}

@misc{FTRanking2026,
  author       = {{Financial Times}},
  title        = {{FT}'s 50 Journals used in {Business} {School} {Research} {Rank} (2026 update)},
  howpublished = {ft.com},
  year         = {2026},
  note         = {Three titles removed (Human Relations, Journal of Business Ethics, Organization Studies); three added (Academy of Management Annals, American Sociological Review, Psychological Science)}
}

\appendix
\section{Sub-discipline assignment of FT50 journals}
\label{app:disc}

The grouping below is anchored in each journal's primary subject area in the Scopus All Science Journal Classification, which is the classification SciVal uses to compute FWCI \citep{Baas2020,ElsevierMetrics2019}. For journals that span multiple ASJC categories the subfield assignments documented for FT50 journals by \citet{Fassin2021} and the Information Systems clustering used by \citet{Vidgen2019} are followed. Practitioner outlets and Innovation Studies are treated as separate categories because their indicator profiles differ systematically from the academic and the broader social-science groups respectively.

\begin{table}[h]
\centering
\caption{Sub-discipline assignment of the 53 FT50 journals.}
\label{tab:disc_mapping}
\small
\begin{tabular}{p{4.4cm}p{8.5cm}}
\toprule
Sub-discipline & Journals \\
\midrule
Economics (5) & American Economic Review; Quarterly Journal of Economics; Journal of Political Economy; Econometrica; Review of Economic Studies \\
Finance (5) & Journal of Finance; Journal of Financial Economics; Review of Financial Studies; Review of Finance; Journal of Financial and Quantitative Analysis \\
Accounting (6) & Accounting Review; Journal of Accounting Research; Journal of Accounting and Economics; Accounting, Organizations and Society; Contemporary Accounting Research; Review of Accounting Studies \\
Management (16) & Academy of Management Journal; Academy of Management Review; Academy of Management Annals; Administrative Science Quarterly; Strategic Management Journal; Journal of Management; Journal of Management Studies; Organization Science; Organization Studies; Human Relations; Human Resource Management; Strategic Entrepreneurship Journal; Entrepreneurship Theory and Practice; Journal of Business Venturing; Journal of Business Ethics; Journal of International Business Studies \\
Marketing (6) & Journal of Marketing; Journal of Marketing Research; Marketing Science; Journal of Consumer Research; Journal of Consumer Psychology; Journal of the Academy of Marketing Science \\
OR / Operations (5) & Management Science; Operations Research; Manufacturing and Service Operations Management; Production and Operations Management; Journal of Operations Management \\
Information Systems (3) & MIS Quarterly: Management Information Systems; Information Systems Research; Journal of Management Information Systems \\
Psychology / OB (3) & Journal of Applied Psychology; Psychological Science; Organizational Behavior and Human Decision Processes \\
Sociology (1) & American Sociological Review \\
Practitioner (2) & Harvard Business Review; MIT Sloan Management Review \\
Innovation Studies (1) & Research Policy \\
\bottomrule
\end{tabular}
\end{table}

\end{document}